\documentclass[a4paper,10pt]{article}
\usepackage[latin1]{inputenc}
\usepackage[english]{babel}
\usepackage{graphicx}
\begin{document}
\title{Isotropisation of Bianchi class A models with a minimally coupled scalar field and a perfect fluid}
\author{Stéphane Fay\\
Laboratoire Univers et Théories(LUTH), CNRS-UMR 8102\\
Observatoire de Paris, F-92195 Meudon Cedex\\
France\\
\small{Steph.Fay@Wanadoo.fr}}
\date{\today}
\maketitle
\begin{abstract}
We look for the necessary conditions allowing the Universe isotropisation in presence of a minimally 
coupled and massive scalar field with a perfect fluid. We conclude that it arises only when the Universe is scalar field dominated, leading to flat spacelike sections and accelerated expansion, and examine the case of a SUGRA theory.\\\\
\end{abstract}
Published in Class. Quantum Grav. vol 21, 6, 1609-1621, 2004
\section{Introduction}\label{s0}
Today the introduction of scalar fields in cosmology obeys major reasons taking roots in the extension of the standard particle physics model such as supersymmetry which requires additional degrees of freedom represented by these fields. In a general way, most of the theories predicting extra dimensions at high energy could generate scalar fields\cite{EllKalOliYok99} via compactification processes. Supergravity theory(SUGRA)\cite{BraMar99A, BraMar99B} related to supersymmetry concept or Higgs mechanism which allows us to explain the mass of particles also imply some scalar fields. From an observational point of view they could be responsible for dark matter\cite{MatGuzUre99,MatUre00,MatGuz01} as well as dark energy\cite{Per99,Rie98,Spe03,Fay03A} although other explanations exist. They could also solve the so-called cosmological constant problem: most of these scalar fields are massive and thus able to mimic a variable cosmological constant.\\
Let us speak about the geometrical context of this paper. There exist nine anisotropic cosmological models classified by Bianchi in 1897. We will be interested in the curved Bianchi class $A$ models since we have studied the spatially flat Bianchi type $I$ model in \cite{Fay01A} and there is no adapted ADM Hamiltonian formulation for the Bianchi class $B$ models. Among the Bianchi class $A$ models, the Bianchi type $IX$ one contains the solutions of the positively curved isotropic FLRW model and the Bianchi type $II$ one characterizes the strong anisotropic phases\cite{Lid96}. Unless we assume a Universe born isotropic and homogeneous, as instance thanks to a quantum principle selecting this type of particular model among all the possible ones, it is legitimate to ask why our Universe is so symmetric. It seems more natural to suppose that it was initially less symmetric and that it asymptotically evolves to an FLRW model. This is one of the reasons why the study of anisotropic models is so important. It allows us to explore the mechanisms responsible for the isotropisation of our Universe and to put some constraints that may be compared to observations on its final isotropic state. Moreover, from the initial state point of view, the oscillatory approach of the singularity by the Bianchi type $IX$ model is generally considered as more generic than the one of the FLRW models and could be shared by the most general inhomogeneous models as conjectured by Belinskij, Khalatnikov and Lifchitz \cite{BelKhaLif82,BelKhaLif70}.\\
Our goal is to find some scalar field properties allowing the Universe to reach isotropy and then the dynamical behaviours of the metric and potential. In \cite{Fay03}, we have shown that isotropisation of curved class $A$ Bianchi models in presence of a massive scalar field but without a perfect fluid always leads to a late times acceleration which is not necessary the case when there is no curvature\cite{Fay01}. In \cite{Fay01A}, we have seen that in presence of a perfect fluid, the isotropisation of the flat Bianchi type $I$ model leads to a decelerated expansion if asymptotically the difference $p_\phi-\rho_\phi$ between the pressure and the density of the scalar field is proportional to the density $\rho$ of the perfect fluid. What happens when we consider both curvature and perfect fluid? Here, we will try to answer this question.\\
To this end, we will use the ADM Hamiltonian formalism\cite{Nar72} to get a first order equations system that we will study by help of dynamical systems analysis\cite{WaiEll97}. Most of times, dynamical analysis of the field equations in cosmology rest on the orthonormal frame formalism and Hubble-normalized variables as shown in Wainwright and Ellis book\cite{WaiEll97}. It allows us to study a large number of cosmological models in various situations, even the most complex one such as the inhomogeneous cosmologies\cite{UggElsWaiEll03} or the presence of magnetic fields\cite{HorWai03}, finding and classifying all the equilibrium points of these systems. Some scalar-tensor theories have also been studied in this way but, to our knowledge, their forms were always completely specified, i.e. they did not contain any unspecified function of the scalar field. Here, we want to consider a class of scalar-tensor theories containing two unspecified functions of the scalar field and just look for the stable isotropic state the Universe can reach. Hence, we aim to study a larger class of scalar-tensor theories than usually and it is one of the reasons why we have not used the powerful orthonormal frame formalism but rather the more traditional Hamiltonian ADM formalism which have proved to be useful in such a case\cite{Fay00A}.\\
The plan of this work is as follows: in the second part we establish the Hamiltonian field equations and, after having remembered the results we obtained without a perfect fluid, we study the isotropisation process when it is present. We discuss the physical meaning of our results in the last section.
\section{Field equations and dynamical analysis}\label{s2}
In the first subsection, we derive the Hamiltonian field equations and in the second one, we use dynamical systems analysis to study the stable isotropic states.
\subsection{Field equations}\label{s20}
We will use the following metric, reflecting the 3+1 decomposition of spacetime:
\begin{equation}\label{metrique}
ds^2 = -(N^2 -N_i N^i )d\Omega^2 + 2N_i d\Omega\omega^i + R_0 ^2 g_{ij}\omega^i \omega^j 
\end{equation}
The $\omega_i$ are the 1-forms generating the Bianchi homogeneous spaces, $N$ and $N_i$ are the lapse and shift functions and $g_{ij}$ are the metric functions parameterised by Misner\cite{Mis69} as: 
\begin{eqnarray*}
g_{11}&=&e^{-2\Omega+\beta_++\sqrt{3}\beta_-}\\
g_{22}&=&e^{-2\Omega+\beta_+-\sqrt{3}\beta_-}\\
g_{33}&=&e^{-2\Omega-2\beta_+}
\end{eqnarray*}
The $\beta_\pm$ functions describe the anisotropy whereas $\Omega$ is the metric isotropic part. The action of the minimally coupled and massive scalar field theory with a non tilted perfect fluid writes:
\begin{equation} \label{action}
S=(16\pi)^{-1}\int \left[R-(3/2+\omega(\phi))\phi^{,\mu}\phi_{,\mu}\phi^{-2} -U(\phi)+16\pi c^4 L_m\right]\sqrt{-g}d^4 x
\end{equation}
$U$ is the potential of the scalar field $\phi$ whose coupling with the metric is described by the Brans-Dicke coupling function $\omega$\footnote{A scalar field transformation sometimes allows to reduce the two unspecified functions $\omega$ and $U$ to a single function. However, the transformation is not always analytically possible and it is why it is more general to consider the two functions.}. $L_m$ is the Lagrangian of the non tilted perfect fluid whose equation of state is $p=(\gamma-1)\rho$ with $\gamma\in\left[1,2\right]$. It describes a dust fluid when $\gamma=1$ and a radiative fluid when $\gamma=4/3$. The other important value is $\gamma=0$ and corresponds to a cosmological constant which has been discussed in \cite{Fay03}. Technical details allowing to get the ADM Hamiltonian from the action (\ref{action}) have been given in \cite{Rya72,MatRyaTot73,Fay01}. Hence we write directly:
\begin{equation} \label{hamiltonien}
H^2 = p_+ ^2 +p_- ^2 +12\frac{p_\phi ^2 \phi^2}{3+2\omega}+24\pi^2 R_0 ^6 e^{-6\Omega}U+\delta e^{3(\gamma-2)\Omega}+V(\Omega,\beta_+,\beta_-)
\end{equation}
with $p_\pm$ and $p_\phi$, respectively the conjugate momenta of the $\beta_\pm$ variables and the scalar field. $V(\Omega,\beta_+,\beta_-)$ is the curvature potential characterising each curved Bianchi class $A$ model and given in table \ref{tab0}.
\begin{table}[h]	
\begin{center}
\begin{tabular}{|l|l|}
\hline
Type&Expression of $V(\Omega,\beta_+,\beta_-)$\\
\hline
$II$&$24\pi^2R_0^4e^{-4\Omega+4\beta_++\sqrt{3}\beta_-}$\\
$VI_0$&$24\pi^2R_0^4e^{-4\Omega+4\beta_+}(\cosh{4\sqrt{3}\beta_-}+1)$\\
$VII_0$&$24\pi^2R_0^4e^{-4\Omega+4\beta_+}(\cosh{4\sqrt{3}\beta_-}-1)$\\
$VIII$&$24\pi^2R_0^4e^{-4\Omega}\mbox{[}e^{4\beta_+}(\cosh{4\sqrt{3}\beta_-}-1)+
1/2e^{-8\beta_+}+2e^{-2\beta_+}\cosh{-2\sqrt{3}\beta_-}\mbox{]}$\\
$IX$&$24\pi^2R_0^4e^{-4\Omega}\mbox{[}e^{4\beta_+}(\cosh{4\sqrt{3}\beta_-}-1)+
1/2e^{-8\beta_+}-2e^{-2\beta_+}\cosh{-2\sqrt{3}\beta_-}\mbox{]}$\\
\hline
\end{tabular}
\end{center}
\caption{\label{tab0}Curvature potentials for Bianchi type $II$, $VI_0$, $VII_0$, $VIII$ and $IX$ models}
\end{table}
$\delta$ is a positive constant proportional to $(\gamma-1)\rho_0$. Using (\ref{hamiltonien}), the Hamiltonian equations are:
\begin{equation} \label{beta}
\dot{\beta}_ \pm = \frac{\partial H}{\partial p_ \pm}=\frac{p_\pm}{H}
\end{equation}
\begin{equation} \label{phi}
\dot{\phi}=\frac{\partial H}{\partial p_\phi}=\frac{12\phi^2 p_\phi }{(3+2\omega)H}
\end{equation}
\begin{equation} \label{p}
\dot{p}_\pm=-\frac{\partial H}{\partial \beta_ \pm}=-\frac{\partial V}{2H\partial \beta_ \pm}
\end{equation}
\begin{equation} \label{pphi}
\dot{p}_\phi=-\frac{\partial H}{\partial \phi}=-12\frac{\phi p_\phi ^2}{(3+2\omega)H}+12\frac{\omega_\phi \phi^2 p_\phi ^2 }{(3+2\omega)^2 H}-12\pi^2 R_0 ^6 \frac{e^{-6\Omega}U_\phi }{H}
\end{equation}
\begin{equation} \label{H}
\dot{H}=\frac{dH}{d\Omega}=\frac{\partial H}{\partial \Omega}=-72\pi^2 R_0 ^6 \frac{e^{-6\Omega}U}{H}+3/2\delta(\gamma-2)\frac{e^{3(\gamma-2)\Omega}}{H}+\frac{\partial V}{2H\partial\Omega}
\end{equation}
In this paper, we will choose $N_i=0$, i.e. a diagonal metric, and we derive\cite{Nar72} that 
$$
N=\frac{12\pi R_0^3e^{-3\Omega}}{H}$$
Now, we have to rewrite these equations with some variables, bounded in the neighbourhood of the isotropy. In \cite{Fay03} we had used the following variables common to all the curved Bianchi class $A$ models:
\begin{equation} \label{v1}
x_\pm=p_\pm H^{-1}
\end{equation}
\begin{equation} \label{v2}
y=\pi R_0^3\sqrt{U}e^{-3\Omega}H^{-1}
\end{equation}
\begin{equation} \label{v3}
z=p_{\phi}\phi(3+2\omega)^{-1/2}H^{-1}
\end{equation}
and $\phi$ the scalar field. These variables are real as long as $U>0$ and $3+2\omega>0$ which is necessary to respect the weak equivalence principle. Each of them has a physical interpretation:
\begin{itemize}
\item $x_\pm^2$ are proportional to the shear parameters $\Sigma_\pm$ defined in \cite{WaiEll97}.
\item $y^2$ is proportional to $(\rho_\phi-p_\phi)/(d\Omega/dt)^2$, $(d\Omega/dt)^2$ being the Hubble variable when the Universe is isotropic, $\rho_\phi$ and $p_\phi$ the density and pressure of the scalar field.
\item $z^2$ is proportional to $(\rho_\phi+_\phi)/(d\Omega/dt)^2$, $(d\Omega/dt)^2$.
\item We deduce from these two last points that the density parameter $\Omega_\phi$ for the scalar field is a linear combination of $y^2$ and $z^2$ or, when the scalar field is quintessent, that these two variables are proportional to $\Omega_\phi$.
\end{itemize}
We had also defined some "w" variables characterising the curvature of each Bianchi model and which are shown in the table \ref{tab1}. They are related to the three $N_i$ variables describing the curvature in the paper of Horwood and Wainwright\cite{RosJan88} or in the book edited by Wainwright and Ellis\cite{WaiEll97} and defined by using a symmetry group structure. In this last book, the curvature of the Bianchi type $II$ model, $VI_0$ and $VII_0$ models, $VIII$ and $IX$ models are respectively described by $N_1$ , $(N_2,N_3)$ and $(N_1,N_2,N_3)$ variables. Here, in a similar way, we could redefine three variables $w_i$, $i=1,2,3$ such that for the Bianchi type $II$ model, $VI_0$ and $VII_0$ models, $VIII$ and $IX$ models, the curvature be described by $(w_1=w)$, $(w_1=w_+, w_2=w_-)$ and $(w_1=w_p w_-,w_2=w_p/w_-,w_3=w_m)$, thus recovering the same unified picture as in \cite{WaiEll97}.\\
\begin{table}[h]
\begin{center}
\begin{tabular}{|c|c|}
\hline
Bianchi models&Associated variables\\
\hline
$II$&$w=\pi R_0^2 e^{-2\Omega+2(\beta_++\sqrt{3}\beta_-)}H^{-1}$\\
\hline
$VI_0$ and $VII_0$&$w_\pm=\pi R_0^2 e^{-2\Omega+2(\beta_+\pm\sqrt{3}\beta_-)}H^{-1}$\\
\hline
$VIII$ and $IX$&$w_p=\pi R_0^2 e^{-2\Omega+2\beta_+}H^{-1}$\\
&$w_m=\pi R_0^2 e^{-2\Omega-2\beta_+}H^{-1}$\\
&$w_-=e^{2\sqrt{3}\beta_-}$\\
\hline
\end{tabular}
\end{center}
\caption{\label{tab1}$w$ variables characterising he curvature of each Bianchi model.}
\end{table}
In this paper, we will also consider an additional variable called $k$ and related to the presence of a perfect fluid. It is defined by
\begin{eqnarray}\label{}
k^2&=&\delta e^{3(\gamma-2)\Omega}H^{-2}\nonumber\\
k^2&=&\delta y^2V^{-\gamma}U^{-1}\label{k}\\\nonumber
\end{eqnarray}
$k$ is proportional to the density parameter of the perfect fluid, one of the main parameters in cosmology. It can be shown by checking that $k^2\propto V^{-\gamma}/(\frac{d\Omega}{dt})^2$. $k$ is not independent from the other variables and when no perfect fluid is considered, $k=0$ strictly.\\
For each Bianchi model, we have rewritten the Hamiltonian constraint and the field equations with these variables in the second appendix.
\subsection{Isotropisation}\label{s21}
In the first subsection, we define the different ways to reach a stable isotropic state. In the second one, we recall our results obtained without the perfect fluid. In the third one, we discuss about their stability. In the fourth one, we extend them by considering the presence of a perfect fluid.
\subsubsection{Different kinds of isotropisation}\label{s21A}
In \cite{Fay03} when no perfect fluid is present, we had defined the isotropy as the convergence of the metric functions to a common form such as the Hubble parameter is the same in any directions. It implied $d\beta_\pm/dt\rightarrow 0$ and $\beta_\pm\rightarrow const$ and thus that it should arise when $p_\pm e^{3\Omega}\rightarrow 0$. This definition is unchanged in presence of a perfect fluid.\\
Different kind of isotropisation may exist, leading to a forever expanding model, a singularity or a static Universe. We had shown that when there is no perfect fluid, isotropy only occurs when $\Omega\rightarrow -\infty$ and $x\rightarrow 0$, i.e. for a forever expanding Universe. Looking at the field equations, we find three ways to reach an isotropic stable state that we have classified in three classes:
\begin{enumerate}
\item Class 1: all the variables but not necessarily the scalar field reach equilibrium with $y\not =0$.
\item Class 2: all the variables but not necessarily the scalar field reach equilibrium with $y =0$.
\item Class 3: all the variables do not reach equilibrium but $x_\pm$ which, as the $w$ functions, have to vanish. 
\end{enumerate}
For the class 2, generally nothing can be deduced about the asymptotic behaviours of the metric functions and potential. It has been numerically observed in a paper in preparation where a non minimally coupling between a scalar field and a perfect fluid is considered. For the class 3, $y$ and $z$ do not necessarily need to reach equilibrium when the Universe isotropises. They just have to be bounded when $\Omega\rightarrow -\infty$, implying that they oscillate. Hence, the signs of their derivatives, which do not asymptotically vanish\footnote{We are assuming that they do not reach equilibrium!}, change continuously\footnote{The variables are bounded}. We have numerically observed the class 3 isotropisation in presence of several scalar fields\cite{FayLum03} and it seems to be associated to an oscillating behaviour of $\ell$.\\
In this paper, we will study the class 1 isotropisation. In the two next subsections, we briefly recall the results we obtained in \cite{Fay03} without a perfect fluid and then discuss the assumptions we have made related to their stability.
\subsubsection{Without the perfect fluid}\label{s211}
The results we obtained in \cite{Fay03} are the following. We define the function $\ell$ of the scalar field:
$$
\ell=\phi U_\phi U^{-1}(3+2\omega)^{-1/2}
$$
The equilibrium points corresponding to class 1 isotropisation are given by $(x_\pm,y,z)=(0,\pm\sqrt{3-\ell^2}(6\sqrt{2})^{-1},\ell/6)$, the $w$ variables related to the curvature (see table \ref{tab0}) being zero. Our conclusion about isotropisation, valid whatever the curved Bianchi class $A$ models when no perfect fluid is present, was that it occurs for a forever expanding Universe ($\Omega\rightarrow -\infty$) and it requires $\ell^2$ to tend to a constant\footnote{It is necessary that $\ell$ tends to a constant otherwise, $\dot z$ as instance, which tends to $\dot\ell/6$, could not be asymptotically  vanishing.} $\ell_0$ smaller than $1$. Then the metric functions tend to $t^{\ell_0^{-2}}$ if $\ell_0\not=0$ or to a De Sitter model otherwise. The Universe is thus asymptotically accelerated and flat. The scalar field asymptotical behaviour may be determined by the asymptotical solution of the first degree differential equation $\dot\phi=2\phi^2 U_\phi(3+2\omega)^{-1}U^{-1}$.\\
For Bianchi type $II$, $VI_0$ and $VII_0$ model, isotropisation will occur at late times if the Hamiltonian $H$ is initially positive and at early times otherwise. It is easily shown by noting that $H$ is a monotonic function of $\Omega$ with a constant sign. Then, using the relation $dt=-Nd\Omega$, it comes that $\Omega$ is a decreasing(increasing) function of the proper time $t$ when $H$ is positive (negative). Since the Universe only isotropises in $\Omega\rightarrow -\infty$, it thus corresponds to late times and forever expanding Universe. For the Bianchi types $VIII$ and $IX$ models, it is not possible to show that $H$ is a monotonic function and thus, the isotropisation time is undetermined.
\subsubsection{Stability of our results}\label{s21B}
The above results or the ones of the present paper are the determination of the isotropic equilibrium points, some necessary conditions for isotropisation and the asymptotical behaviours of some functions in the neighbourhood of these points. However the asymptotical behaviours are determined by calculating the exact solutions for each equilibrium point and they will be correct only if on one hand $\ell$ and in the other hand the variables $(y,z,w)$ (and $k$ when we will consider a perfect fluid) tend sufficiently fast to their equilibrium values. Otherwise, they will be different. Let us explain why.\\\\
The first kind of instability comes from $\ell$. As instance, when we look for $x_\pm$ asymptotical behaviours, we need to calculate $exp(\int \ell^2 d\Omega)$. As explains in the section \ref{s211}, we have shown in \cite{Fay03} that near the isotropic state, $\ell^2$ tends to a constant $\ell_0$ (vanishing or not). Then, in our calculation, we have assumed that asymptotically when $\Omega\rightarrow -\infty$, $exp(\int \ell^2 d\Omega) \rightarrow exp(\ell_0^2\Omega)$ but this is true only if $\ell^2$ tends sufficiently fast to its constant equilibrium value. As instance, if $\ell^2\rightarrow \ell_0^2+\Omega^{-1/2}$, $\ell^2$ tends to a constant but $exp(\int \ell^2 d\Omega)$ does not tend asymptotically to an exponential because of the quantity $\Omega^{-1/2}$. Hence, our results will be valid as long as the following assumption holds:
\begin{itemize}
\item When $\ell$ tends to a constant $\ell_0$ (vanishing or not) such that $\ell^2\rightarrow \ell^2_0+\delta\ell^2$, $\int(\ell^2_0+\delta\ell^2) d\Omega\rightarrow \ell^2_0\Omega+const$.
\end{itemize}
If the last limit is not true, the asymptotical behaviours for the metric functions (and potential) are different from classical power or exponential laws. This problem could be overcame since our results allow to calculate $\phi(\Omega)$ and thus $\ell(\Omega)$. Hence, it should be easy to generalise them by keeping the $\int \ell^2d\Omega$ term instead of considering that it tends to $\ell^2\Omega$ but then they would not be on a closed form.\\\\
The second kind of instability can not be solved so easily. In the same way, the asymptotical behaviours we have determined will be true only if the variables $(y,z,w,k)$ tend sufficiently fast to their equilibrium values. As instance near isotropy we have $y\rightarrow \pm\sqrt{3-\ell^2}(6\sqrt{2})^{-1}$ and when we integrate the differential equation for $x_\pm$, we assume that $exp(\int y^2 d\Omega)\rightarrow exp(\int (3-\ell^2)/72 d\Omega)$. But once again, this is not exact if $y^2$ tends to its equilibrium value slower than $\Omega^{-1}$ and we have to make the same kind of assumption for $(y,z,w,k)$ as for $\ell$. For partly solve this problem, it would be necessary to consider some small perturbations of the exact solutions but until now we have not succeed to get any interesting results, even for the flat model.\\\\
To summarize, the results of this paper related to asymptotical behaviours will be valid for a class 1 isotropisation if the function $\ell$ and the variables $(y,z,w,k)$ tend sufficiently fast to their equilibrium values or, more physically, if the Universe tends sufficiently fast to its isotropic state. The restriction on $\ell$ may be easily solved but the ones on $(y,z,w,k)$ require a more careful examination.  In the following subsection, we consider the isotropisation of a curved Bianchi class A model in presence of a perfect fluid, first when $k$ vanishes and then when it tends to a non vanishing constant.
\subsubsection{With a perfect fluid}\label{s212}
Depending on the vanishing of $k$ near an isotropic equilibrium state, the results summarize in the section \ref{s211} will or will not be modified.\\
\\
\underline{$k\rightarrow 0$}\\
When $k\rightarrow 0$ near isotropy, the results are the same as those found when we consider no perfect fluid. In particular, isotropisation always arise for a forever expanding Universe, i.e. when $\Omega\rightarrow -\infty$. Obviously, we find the same equilibrium points and assuming that $k$ tends sufficiently fast to its equilibrium value(see section \ref{s21B}), we also recover the same asymptotical behaviours. However, the limit $k\rightarrow 0$ plays the role of a new constraint. This fact was noted in \cite{FayLum03} for the flat Bianchi type $I$ model. In this last paper we had shown that the interval of $\ell$ allowing for isotropy was smaller when we consider a perfect fluid such that $k\rightarrow 0$ than without it: in this last case isotropy requires $\ell^2<3$, otherwise $\ell^2<3/2\gamma$\footnote{This inequality rests on the asymptotical behaviour of $k$ and, as discussed in section \ref{s21B}, it may vary if $k$ does no tend sufficiently fast to its equilibrium value. However, the limit $\ell^2<3$ have always to be respected since it is required for the existence of the equilibrium points, independently on how fast the isotropic state is reached.}.\\
Does the limit $k\rightarrow 0$ also change the necessary conditions for isotropy when we consider some curvature? Near equilibrium, the $w$ variables have to vanish and are proportional to $e^{-2\Omega}H^{-1}$. But $e^{-2\Omega}$ diverges and thus, since $w\rightarrow 0$, $H$ have to be larger than $e^{-2\Omega}$, i.e.:
$$
H>>e^{-2\Omega}
$$
Moreover, we have $y^2=Ue^{-2\Omega}e^{-4\Omega}H^{-2}$ and since near isotropy $y$ tends to a non vanishing constant whereas $e^{-4\Omega}H^{-2}$ tends to vanish, we deduce that 
$$
U>>e^{2\Omega}>>V^{-\gamma}
$$
and then from (\ref{k}) that $k\rightarrow 0$. Consequently, starting from the fact that $w\rightarrow 0$, we conclude that $k\rightarrow 0$ without any modification of the necessary condition for isotropisation on the contrary from the Bianchi type $I$ model. Moreover, it means that the energy density $\rho_\phi$ of the scalar field and its pressure $p_\phi$ are such that $U\propto p_\phi-\rho_\phi>>\rho_m$: the Universe is dynamically dominated by the scalar field. Thus, the results obtained in the vacuum (i.e. $k=0$ strictly) are not changed when we consider a perfect fluid such as $k\rightarrow 0$.\\
\\
\underline{$k\not\rightarrow 0$}\\
Now, we consider what happens when $k\not\rightarrow 0$. The necessary condition for isotropy is still $p_\pm e^{3\Omega}\rightarrow 0$ and we have to determine if it occurs for a forever expanding, contracting or static Universe.
\begin{itemize}
\item If it arises for a diverging $\Omega$, it means that at equilibrium, we must have $x_\pm\rightarrow 0$ as explained in the section \ref{s21A}.
\item If it arises for a finite value of $\Omega$, then we must have $p_\pm\rightarrow 0$. 
\begin{itemize}
\item Let us assume that in the same time $x_\pm\not\rightarrow 0$. Since $p_\pm\rightarrow 0$, from (\ref{v1}) we deduce that $H$ have to vanish otherwise $x_\pm\rightarrow 0$. But then $k^2$, which is proportional to the perfect fluid density parameter, diverges and the constraint is not respected because, near isotropy, all the variables have to be bounded as shown in \cite{Fay03}. Hence, $H$ can not tend to zero and $x_\pm$ must vanish near equilibrium.
\item In the same way, when $\Omega$ tends to a non vanishing constant, $H$ can not diverge because then $k\rightarrow 0$ which is not in agreement with the assumption of this subsection.
\end{itemize}
Consequently, when the isotropy occurs for a finite $\Omega$, the Hamiltonian have to tend to a bounded and non vanishing quantity and it is thus the same for the $w$ variables.
\end{itemize}
To summarize, in the neighbourhood of the isotropic state:
\begin{itemize}
\item If $\Omega$ diverges, the equilibrium points are such that $x_\pm\rightarrow 0$.
\item If $\Omega\rightarrow const\not =0$, the equilibrium points are such that $x_\pm\rightarrow 0$ and $w$ variables are non vanishing and bounded.
\end{itemize}
Whatever the curved Bianchi models, the only equilibrium points corresponding to these requirements when solving the field equations\footnote{For the Bianchi type $VIII$ and $IX$ models where the equations are far from being simple, it is not possible to solve them directly. We proceed by putting $x_\pm=0$ and $w_-=1$ in the equations for $x_\pm$, these values being these required for isotropy. We then show that $x_\pm$ can reach equilibrium only if $w_p$ and $w_m$ vanish which allow us to determine the equilibrium values for the other variables. Since $w_p$ and $w_m$ tend to vanish, the Hamiltonian constraint shows that all the variables are bounded. For the other Bianchi models, we can show in the same way as in \cite{Fay03} that all the variables are bounded near the isotropic equilibrium state.} are defined by $(x_\pm,y,z)=(0,\pm\frac{\sqrt{\gamma(2-\gamma)}}{4\sqrt{2}\pi R_0^3\ell},\frac{\gamma}{4\ell})$, the $w$ variables related to the curvature(see table \ref{tab1}) being 0\footnote{There exist some other equilibrium points for which $k$ or $\ell$ may be chosen such that $x_\pm=0$ and the constraint be respected but they correspond to complex values of some variables.}. The Hamiltonian constraint implies that $k^2=1-\frac{3\gamma}{2\ell^2}$ and consequently $\ell^2>\frac{3}{2}\gamma$. This inequality is independent from any assumption on how far the isotropic state is reached. As the $w$ variables are vanishing, it follows that $\Omega$ must diverge and not tend to a constant. Consequently, we calculate that asymptotically the Hamiltonian, whose form is given in the second appendix as a function of the variables, behaves as:
$$
H\rightarrow e^{-\frac{3}{2}(2-\gamma)\Omega}
$$
This is in agreement with the limit $k^2\rightarrow const\not =0$ and the definition for $k$ which also implies that $U\propto V^{-\gamma}$. Hence, the scalar field plays the same dynamical role as the perfect fluid and we can show that their energy densities scales in the same way, preventing any accelerated expansion. We also get that the $w$ variables(but $w_-$ which tends to a non vanishing constant for the Bianchi type $VIII$ and $IX$ variables) all behave as:
$$
w\rightarrow e^{(1-\frac{3\gamma}{2})\Omega}
$$ 
For the considered range of $\gamma$, we derive that $w\rightarrow 0$ only if $\Omega\rightarrow +\infty$. But for the $x_\pm$ variables, it comes:
$$
x_\pm\rightarrow x_0e^{(2-3\gamma)\Omega}(e^{(1+\frac{3\gamma}{2})\Omega}+x_1)
$$
$x_0$ being an integration constant. It follows that if $\gamma\in\left[1,2\right]$ and $\Omega\rightarrow +\infty$, $x$ diverges. Consequently, the isotropic state can not be reached for this range of $\gamma$. Knowing $x_\pm$ and $H$, we calculate that:
$$
p_\pm\rightarrow e^{-\frac{1}{2}(2+3\gamma)\Omega}+cte
$$
Hence, $p_\pm e^{3\Omega}$, the $w$ and $x_\pm$ variables vanish only if $\gamma<2/3$ and $\Omega\rightarrow -\infty$. Then, we find that $e^{-\Omega}\rightarrow t^{\frac{2}{3\gamma}}$ and, from the definition of $y$ and the property $U\propto V^{-\gamma}$, we derive that $U\rightarrow t^{-2}$. This restriction on $\gamma$ does not exist for the flat Bianchi type $I$ model\cite{Fay01A,FayLum03} and does not fit an ordinary perfect fluid such that $\gamma\in\left[1,2\right]$.
\section{Conclusion}\label{s21}
In this work, we have determined the necessary but not sufficient conditions for class 1 isotropisation of curved Bianchi class $A$ models when a minimally and massive scalar field with a perfect fluid are considered. We have assumed that $U>0$ and $3+2\omega>0$ such that the weak energy principle is respected. Moreover, some of our results related to asymptotical behaviours are valid as long as the isotropic state is reached sufficiently fast.\\

\begin{figure}[h]
\centering
\includegraphics[width=\textwidth]{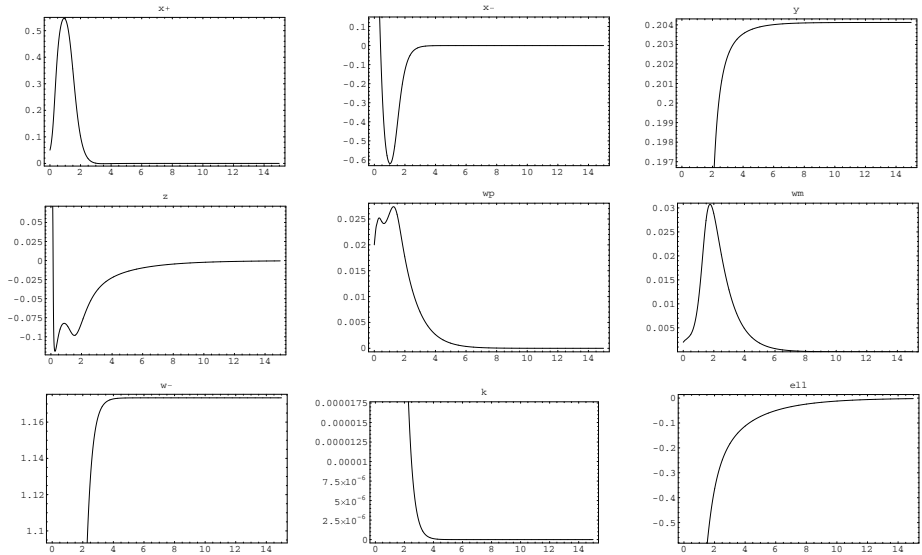}
\caption{\scriptsize{\label{fig1}Isotropisation of SUGRA theory for Bianchi type $IX$ model. Initial conditions and potential parameters respectively are $(x_+,x_-,y,z,w_p,w_m,w_-,\phi)=(0.05,0.83,0.025,0.12,0.02,0.002,0.2,0.14)$ and $(\Lambda,m,n)=(2.0,1.1,0.15)$.}}
\end{figure}

We can distinguish two cases depending on the vanishing of $k$, a variable proportional to the perfect fluid density parameter $\Omega_m$. When isotropy occurs with $k\rightarrow 0$, we have thus $\Omega_m\rightarrow 0$, $U\propto p_\phi-\rho_\phi>\rho_m$ and the results are the same as in $\cite{Fay03}$ where no perfect fluid is present:\\
\\
\emph{\underline{Class 1 isotropisation with $\Omega_m\rightarrow 0$}:\\
A necessary condition for isotropisation of curved Bianchi class $A$ models in presence of a minimally and massive scalar field such that $\Omega_m\rightarrow 0$ will be that the quantity $\ell=\phi U_\phi U^{-1}(3+2\omega)^{-1/2}$ tends to a constant $\ell_0$, whose square is smaller than one. For the Bianchi type $II$, $VI_0$ and $VII_0$ models, it arises at late (early) times if the Hamiltonian is initially positive(negative). For the Bianchi type $VIII$ and $IX$ models, the time of isotropisation is undetermined. If $\ell_0 \not =0$, the metric functions tend to a power law $t^{\ell_0^{-2}}$ and the potential vanishes as $t^{-2}$. If $\ell_0 =0$, the Universe tend to a De Sitter model and the potential to a constant. The isotropisation process always leads to a flat and accelerated Universe.}\\
\\
Considering the limit near isotropy of the Hamiltonian equation for $\dot{\phi}$ rewritten with the normalised variables (see appendice), we deduce that the scalar field asymptotically behaves as the limit of the solution for 
$$
\dot{\phi}=2\phi^2U_\phi U^{-1}(3+2\omega)^{-1}
$$
as $\Omega\rightarrow -\infty$, in the same way as in \cite{Fay03}. This last equation allows us to deduce the asymptotical behaviour of $\ell(\Omega)$ when we specify $\omega$ and $U$. The second result of this work concerns the case for which $k$, or equivalently $\Omega_m$, tends to a non vanishing constant implying that $U\propto p_\phi-\rho_\phi\propto \rho_m$. We have then:\\
\\
\emph{\underline{Class 1 isotropisation with $\Omega_m\rightarrow const\not =0$}:\\
The isotropisation of curved Bianchi class $A$ models in presence of a minimally and massive scalar field such that $\Omega_m\rightarrow const\not =0$ is impossible if the perfect fluid is an ordinary one such that $\gamma\in\left[1,2\right]$. It will only occur if $\gamma<2/3$, which generally corresponds to a quintessent fluid equation of state.}\\
\\
Now, we examine these results with respect to supergravity. In \cite{BraMar99A, BraMar99B}, it is shown that quintessence theories should be based on supergravity. A scalar tensor theory is then derived, defined by $\omega+3/2=\phi^2$ and $U=\Lambda^{4+m}\phi^{-m}e^{\frac{n}{2}\phi^2}$. It is able to solve the coincidence problem and even the fine tuning problem if $m\geq 11$. What about class 1 isotropisation? We calculate that:
$$
\ell^2=(\frac{n\phi^2-m}{\sqrt{2}\phi})^2
$$
and when no matter is present or if $k\rightarrow 0$, the scalar field asymptotically behaves as:
$$
\phi\rightarrow \pm\sqrt{\frac{m-\phi_0e^{2n\Omega}}{n}}
$$
$\phi_0$ being an integration constant.\\
When $n>0$, $\phi\rightarrow (m/n)^{1/2}$ implying that $m$ should be positive. $\ell\rightarrow 0$ and the necessary conditions for isotropisation are thus respected. If it arises, the Universe tends to a De-Sitter model. It could thus describe the inflationary period, before the domination of the matter. This case is plotted on figure \ref{fig1} for the Bianchi type $IX$ model.\\
When $n<0$, the scalar field behaves as $\phi\rightarrow \pm\sqrt{\frac{-\phi_0e^{2n\Omega}}{n}}$. It is defined when $\Omega\rightarrow -\infty$ if $\phi_0>0$ and then diverges. It follows that $\ell$ also diverges and thus a class 1 isotropisation is not possible as confirmed by numerical simulations.\\
Summarising, if the Universe isotropises, this theory issued from SUGRA leads an anisotropic curved Universe to a flat isotropic De Sitter one dominated by the scalar field. It could be a good description for an inflationary period. Numerical simulations have not shown any class 2 or 3 isotropisation.
\\
\\
In conclusion, we knew that when no perfect fluid is present, the class 1 isotropisation of an anisotropic curved Universe may lead the Universe to flat spacelike sections and accelerated expansion if some necessary conditions are respected. The question was: does this acceleration, due to the presence of curvature, always exist in presence of a perfect fluid. The answer is "yes" when the density parameter of the perfect fluid asymptotically vanishes. Then, its presence does not change the asymptotic isotropic state or the necessary conditions to reach it. Contrary to the flat Bianchi type $I$ model for which an isotropic state such that $\Omega_m\rightarrow const\not =0$ may exist, the perfect fluid and the scalar field playing the same dynamical role, the isotropic state in presence of curvature is always scalar field dominated but if the perfect fluid is an exotic one. Future research should be concerned by a scalar field which violates the energy conditions, i.e. such that $\omega<-3/2$ or $U<0$ or which is not minimally coupled to a perfect fluid.  This last possibility, which would allow to extend our results to the Hyperextended Scalar Tensor theory (i.e. with a varying gravitation function) with a perfect fluid, is currently under consideration in a paper in preparation.
\section*{Acknowledgements}
I thank Mr Jean-Pierre Luminet for useful discussion and carefull reading of the manuscript. I also thank anonymous referees for improving the manuscript.
\appendix
\section{Field equations of the curved Bianchi models with normalised variables}\label{a2}
\underline{Bianchi type $II$}\\
The Hamiltonian constraint writes:
\begin{equation} \label{cII}
x_+^2+x_-^2+24y^2+12z^2+12w^2+k^2=1
\end{equation}
The Hamiltonian equations become:
\begin{equation} \label{eq1II}
\dot{x}_+=72y^2x_++24w^2x_+-24w^2-3/2(\gamma-2)k^2x_+
\end{equation}
\begin{equation} \label{eq2II}
\dot{x}_-=72y^2x_-+24w^2x_- -24\sqrt{3}w^2-3/2(\gamma-2)k^2x_-
\end{equation}
\begin{equation} \label{eq3II}
\dot{y}=y(6\ell z+72y^2-3+24w^2)-3/2(\gamma-2)k^2y
\end{equation}
\begin{equation} \label{eq4II}
\dot{z}=y^2(72z-12\ell)+24w^2z-3/2(\gamma-2)k^2z
\end{equation}
\begin{equation} \label{eq5II}
\dot{w}=2w(x_++\sqrt{3}x_-+12w^2+36y^2-1)-3/2(\gamma-2)k^2w
\end{equation}
To get an autonomous system, we need a first order equation for $\phi$. Rewriting (\ref{phi}), it comes:
\begin{equation}\label{eq6II}
\dot\phi=12\frac{z\phi}{\sqrt{3+2\omega}}
\end{equation}
This equation is the same for any Bianchi models. The equation for $\dot H$ may be rewritten as:
\begin{equation}\label{HII}
\dot{H}=-H(72y^2+24w^2+\frac{3}{2}(\gamma-2)k^2)
\end{equation}

\underline{Bianchi $VI_0$ and $VII_0$ models}\\
The Hamiltonian constraint writes:
\begin{equation}\label{CVI}
x_+^2+x_-^2+24y^2+12z^2+12(w_+\pm w_-)^2+k^2=1
\end{equation}
and the Hamiltonian equations become:
\begin{eqnarray}
&\dot{x}_+=72y^2x_++24(x_+-1)(w_-\pm w_+)^2-3/2(\gamma-2)k^2x_+\label{eq1VI}&\\
&\dot{x}_-=72y^2x_-+24x_-(w_-\pm w_+)^2+24\sqrt{3}(w_-^2-w_+^2)-3/2(\gamma-2)k^2x_-\label{eq2VI}&\\
&\dot{y}=y(6\ell z+72y^2-3+24(w_-\pm w_+)^2)-3/2(\gamma-2)k^2y&\\
&\dot{z}=y^2(72z-12\ell)+24z(w_-\pm w_+)^2-3/2(\gamma-2)k^2z&\\
&\dot{w}_+=2w_+\left[x_++\sqrt{3}x_-+12(w_-\pm w_+)^2+36y^2-1\right]-3/2(\gamma-2)k^2w_+\label{eq52}&\\
&\dot{w}_-=2w_-\left[x_+-\sqrt{3}x_-+12(w_-\pm w_+)^2+36y^2-1\right]-3/2(\gamma-2)k^2w_-\label{eq62}&\\\nonumber
\end{eqnarray}
The equation for $\dot H$ is:
\begin{equation}\label{HVI}
\dot{H}=-H\left[72y^2+24(w_+\pm w_-)^2+\frac{3}{2}(\gamma-2)k^2\right]
\end{equation}

\underline{Bianchi $VIII$ and $IX$ models}\\
The Hamiltonian constraint writes:
\begin{eqnarray*}
&x_+^2+x_-^2+24y^2+12z^2+12\mbox[w_p^3(1+w_-^4)\pm 2 w_-(w_mw_p)^{3/2}(1+w_-^2)+&\\
&w_-^2(w_m^3-2w_p^3)\mbox](w_-^2w_p)^{-1}+k^2=1&\\
\end{eqnarray*}
and the Hamiltonian equations are:
\begin{eqnarray}
&\dot{x}_+=72y^2x_++24\{w_p^3(x_+-1)(1+w_-^4)\pm w_-(1+2x_+)(w_mw_p)^{3/2}(1+w_-^2)&\nonumber\\
&+w_-^2\left[(2+x_+)w_m^3-2(x_+-1)w_p^3\right]\}(w_-^2w_p)^{-1}-3/2(\gamma-2)k^2x_+\label{eq13}&\\
&\dot{x}_-=72y^2x_-+24\{w_p^3\left[w_-^4(x_--\sqrt{3})+x_-+\sqrt{3})\right]\pm w_-(w_mw_p)^{3/2}\mbox{[}w_-^2&\nonumber\\
&(-\sqrt{3}+2x_-)+(\sqrt{3}+2x_-)\mbox{]}+w_-^2x_-(w_m^3-2w_p^3)\}(w_-^2w_p)^{-1}\label{eq23}&\\
&-3/2(\gamma-2)k^2x_-&\nonumber\\
&\dot{y}=y\{6\ell z+72y^2-3+24\mbox{[}w_p^3(1+w_-^4)\pm 2(w_mw_p)^{3/2}w_-(1+w_-^2)+&\nonumber\\
&w_-^2(w_m^3-2w_p^3)\mbox{]}(w_-^2w_p)^{-1}\}-3/2(\gamma-2)k^2y&\\
&\dot{z}=y^2(72z-12\ell)+24z\mbox{[}w_p^3(1+w_-^4)\pm 2(w_mw_p)^{3/2}w_-(1+w_-^2)+&\nonumber\\
&w_-^2(w_m^3-2w_p^3)\mbox{]}(w_-^2w_p)^{-1}-3/2(\gamma-2)k^2z&\label{eq43}\\
&\dot{w}_p=w_p\{-2+2x_++72y^2+24\mbox{[}w_p^3(1+w_-^4)\pm2w_-(w_mw_p)^{3/2}(1+w_-^2)&\nonumber\\
&+w_-^2(w_m^3-2w_p^3)\mbox{]}(w_-^{2}w_p)^{-1}\}-3/2(\gamma-2)k^2w_p&\label{eq53}\\
&\dot{w}_m=w_m\{-2-2x_++72y^2+24\mbox{[}w_p^3(1+w_-^4)\pm2w_-(w_mw_p)^{3/2}(1+w_-^2)&\nonumber\\
&+w_-^2(w_m^3-2w_p^3)\mbox{]}(w_-^{2}w_p)^{-1}\}-3/2(\gamma-2)k^2w_m&\label{eq63}\\
&\dot{w}_-=2\sqrt{3}w_-x_-&\label{eq73}\\\nonumber
\end{eqnarray}
and
\begin{eqnarray}\label{hamp3}
&\dot{H}=-H\mbox{[}72y^2+24(\pm 2\frac{w_p^{1/2}w_m^{3/2}}{w_-}\pm 2w_p^{1/2}w_m^{3/2}w_--2w_p^2+\frac{w_p^2}{w_-^2}+&\nonumber\\
&w_p^2w_-^2+\frac{w_m^3}{w_p})+\frac{3}{2}(\gamma-2)k^2\mbox{]}&\\\nonumber
\end{eqnarray}
\bibliographystyle{unsrt}

\begin{thebibliography}{10}

\bibitem{EllKalOliYok99}
John Ellis, Nemanja Kaloper, Keith~A. Olive, and Jun'ichi Yokoyama.
\newblock Topological ${R}^4$ inflation.
\newblock {\em Phys. Rev.}, D59:103503, 1998.

\bibitem{BraMar99A}
Ph. Brax and J.~Martin.
\newblock Quintessence and supergravity.
\newblock {\em Phys.Lett.}, B468:40--45, 1999.

\bibitem{BraMar99B}
Ph. Brax and J.~Martin.
\newblock The robustness of quintessence.
\newblock {\em Phys.Rev.}, D61:103502, 2000.

\bibitem{MatGuzUre99}
Tonatiuh Matos, Francisco~S. Guzm{\'a}n, and L.~Arturo Une{\~n}a-L{\'o}pez. \newblock Scalar field as dark matter in the {U}niverse.
\newblock {\em Class.Quant.Grav.}, 17:1707--1712, 1999.

\bibitem{MatUre00}
Tonatiuh Matos and L.~Arturo~Ure\ na~L\`opez.
\newblock Quintessence and scalar field matter in the {U}niverse.
\newblock {\em Class.Quant.Grav.}, 17:L75--L81, 2000.

\bibitem{MatGuz01}
Tonatiuh Matos and F.~Siddhartha Guzm{\'a}n.
\newblock On the space time of a galaxy.
\newblock {\em Class.Quant.Grav.}, 18:5055--5064, 2001.

\bibitem{Fay03A}
S.~Fay.
\newblock Scalar fields properties for flat galactic rotation curves.
\newblock {\em Astronomy and Astrophysics}, 413:799, 2004.

\bibitem{Per99}
S.~Perlmutter et~al.
\newblock Measurements of ${\Omega}$ and ${\Lambda}$ from 42 {H}ight-{R}edshift
  {S}upernovae.
\newblock {\em Astrophysical Journal}, 517:565--586, 1999.

\bibitem{Rie98}
Adam Riess et~al.
\newblock Observational evidence from {S}upernovae for an accelerating
  {U}niverse and a cosmological constant.
\newblock {\em Astrophysical Journal}, 116:1009, 1998.

\bibitem{Spe03}
D.N. Spergel et~al.
\newblock First year wilkinson microwave anisotropy probe (wmap) observations:
  Determination of cosmological parameters.
\newblock {\em Submitted to Astrophys. J.}, 2003.

\bibitem{Fay01A}
S.~Fay.
\newblock Isotropisation of the minimally coupled scalar-tensor theory with a
  massive scalar field and a perfect fluid in the {B}ianchi type {I} model.
\newblock {\em Class. Quantum Grav}, 19, 2:269--278, 2002.

\bibitem{Lid96}
James~E. Lidsey.
\newblock Symetric vacuum scalar-tensor cosmology.
\newblock {\em Class. Quantum Grav.}, 13:2449, 1996.

\bibitem{BelKhaLif82}
V.~A. Belinskii, I.~M. Khalatnikov, and E.~M. Lifshitz.
\newblock A general solution of the {E}instein equations with a time
  singularity.
\newblock {\em Advances in Physics}, 31, 6:639--667, 1982.

\bibitem{BelKhaLif70}
V.~A. Belinskii, I.~M. Khalatnikov, and E.~M. Lifshitz.
\newblock Oscillatory approach to a singularity in the relativistic cosmology.
\newblock {\em Advances in Physics}, 19:525--573, 1970.

\bibitem{Fay03}
S.~Fay.
\newblock Isotropisation of {B}ianchi class {A} models with curvature for a
  minimally coupled scalar tensor theory.
\newblock {\em Class. Quantum Grav}, 20, 7, 2003.

\bibitem{Fay01}
S.~Fay.
\newblock Isotropisation of {G}eneralised-{S}calar {T}ensor theory plus a
  massive scalar field in the {B}ianchi type {I} model.
\newblock {\em Class. Quantum Grav}, 18:2887--2894, 2001.

\bibitem{Nar72}
Hidekazu Nariai.
\newblock Hamiltonian approach to the dynamics of {E}xpanding {H}omogeneous
  {U}niverse in the {B}rans-{D}icke cosmology.
\newblock {\em Prog. of Theo. Phys.}, 47,6:1824, 1972.

\bibitem{WaiEll97}
J.~Wainwright and G.F.R. Ellis, editors.
\newblock {\em Dynamical Systems in Cosmology}.
\newblock Cambridge University Press, 1997.

\bibitem{UggElsWaiEll03}
Claes Uggla, Henk van Elst, John Wainwright, and George F.~R. Ellis.
\newblock The past atttractor in inhomogeneous cosmology.
\newblock {\em gr-qc/0304002, submitted for publication to Physical Review D},
  2003.

\bibitem{HorWai03}
Joshua~T. Horwood and John Wainwright.
\newblock Asymptotic regimes of magnetic {B}ianchi cosmologies.
\newblock {\em submitted to Gen. Rel. Grav}, 2003.

\bibitem{Fay00A}
S.~Fay.
\newblock Hamiltonian study of the generalized scalar-tensor theory with
  potential in a {B}ianchi type {I} model.
\newblock {\em Class. Quantum Grav.}, 17:891--902, 2000.

\bibitem{Mis69}
C.~W. Misner.
\newblock Mixmaster {U}niverse.
\newblock {\em Phys. Rev. Lett.}, 22:1071, 1969.

\bibitem{Rya72}
Michael~P. Ryan.
\newblock {\em Hamiltonian cosmology}.
\newblock Springer-Verlag, 1972.

\bibitem{MatRyaTot73}
R.~A. Matzner, M.~P. Ryan, and E.~T. Toton.
\newblock The {B}rans-{D}icke theory and anisotropic cosmologies.
\newblock {\em Nuovo Cim.}, 14B:161, 1973.

\bibitem{RosJan88}
Kjell Rosquist and Robert~T. Jantzen.
\newblock Unified regularisation of bianchi cosmology.
\newblock {\em Phys. Rep.}, 166:90--124, 1988.

\bibitem{FayLum03}
S.~Fay and J.~P. Luminet.
\newblock Isotropisation of flat homogeneous cosmologies in presence of
  minimally coupled massive scalar fields with a perfect fluid.
\newblock {\em Accepted in Class. Quantum Grav.}, 2004.

\end{thebibliography}

\end{document}